\documentclass[aps,prd,twocolumn,tightenlines,showpacs,amsmath,amssymb,nofootinbib]{revtex4-1}

\usepackage{graphicx}
\usepackage{amsmath,amssymb}
\usepackage{bm}
\usepackage{epsfig}
\usepackage{color}              
\usepackage{hyperref}

\def\be{\begin{equation}}
\def\ee{\end{equation}}
\def\ba{\begin{eqnarray}}
\def\ea{\end{eqnarray}}
\def\bal{\begin{align}}
\def\eal{\end{align}}
\def\bald{\begin{aligned}}
\def\eald{\end{aligned}}
\def\nn{\nonumber}
\newcommand{\per}{\, .}
\newcommand{\com}{\, ,}
\newcommand{\eref}[1]{Eq.~(\ref{#1})}

\newcommand{\uzua}{\rm{U(1)}_{\scriptscriptstyle \cal Z} \times \rm{U(1)}_{\scriptscriptstyle \cal A}}
\newcommand{\ua}{\rm{U(1)}_{\scriptscriptstyle \cal A}}
\newcommand{\uz}{\rm{U(1)}_{\scriptscriptstyle \cal Z}}
\newcommand{\Eir}{\mathcal{E}_{\rm \scriptscriptstyle IR}}

\begin{document}

\title{Diamagnetic Vortices in Chern Simons Theory
}

\date{\today}

\author{Mohamed M. Anber}
\email[]{mohamed.anber@epfl.ch}
\author{Yannis Burnier}
\email[]{yannis.burnier@epfl.ch}
\author{Eray Sabancilar} 
\email[]{eray.sabancilar@epfl.ch}
\author{Mikhail Shaposhnikov}
\email[]{mikhail.shaposhnikov@epfl.ch}
\affiliation{ 
Institut de Th\'eorie des Ph\'enom\`enes Physiques, Ecole Polytechnique F\'ed\'erale de Lausanne, 
CH-1015 Lausanne, Switzerland.
}

\begin{abstract}
We find a new type of topological vortex solution in the $\uzua$ Chern Simons gauge theory in the presence of a $\ua$ magnetic field background. In this theory $\uz$ is broken spontaneously by the $\ua$ magnetic field. These vortices exhibit long range interactions as they are charged under the unbroken $\ua$. They deplete the $\ua$ magnetic field near their core and also break both $C$ and $P$ symmetries. Understanding the nature of these vortices sheds light on the ground state structure of the superconductivity studied in \cite{3dsuperconductor}. We also study the Berezinsky-Kosterlitz-Thouless phase transition in this class of theories and point out that superconductivity can be achieved at high temperatures by increasing the $\ua$ magnetic field.
\end{abstract}

\maketitle


\section{Introduction}

Chern Simons gauge theories \cite{Deser:1981wh,Deser:1982vy} have a wide class of interesting applications in both high energy and condensed matter physics. In the former case, they appear upon dimensionally reducing finite temperature four dimensional field theories with fermions \cite{Redlich:1984md}. In condensed matter physics, they are used to model the quantum Hall effect \cite{Frohlich:1990xz,Frohlich:1991wb,PhysRevB.80.205319}. 

In an accompanying paper, we show that a $2+1$ dimensional $\uzua$ Chern Simons theory exhibits superconductivity to arbitrarily high temperatures. This is achieved by turning on a constant $\ua$ magnetic field that spontaneously breaks the $\uz$ symmetry \cite{3dsuperconductor}. It is well known that the vacuum manifold of spontaneously broken gauge symmetries can have non-trivial topology \cite{Kibble:1976sj}. In particular, breaking ${\rm U(1)} \to 1$ gives rise to topological vortices \cite{Abrikosov:1956sx,Nielsen:1973cs}. This work is devoted to the study of the non-perturbative sector of the $\uzua$ theory in a constant $\ua$ magnetic field background, which is crucial to understand the nature of the superconducting vacuum at high temperatures. In fact, at low temperatures these vortices cannot alter the vacuum structure since it is expensive to produce them. However, at high enough temperatures they may proliferate and change the properties of the superconducting vacuum. This is the celebrated Berezinsky-Kosterlitz-Thouless (BKT) transition \cite{Berezinsky:1970fr,Kosterlitz:1973xp} that occurs in 2 dimensional systems. 

The vortices that we describe in this work share the main characteristics of the vortices that we have studied in a previous work \cite{Anber:2015kxa}. The most striking property of both vortices is that they exhibit long range interactions since $\ua$ remains unbroken in the infrared (see our accompanying work \cite{3dsuperconductor} and also \cite{Anber:2015kxa} for a detailed discussion). Therefore, a system of a vortex and an antivortex will be logarithmically confined. As we mentioned, the vortices in this work exist in a constant $\ua$ magnetic field background. Interestingly, we find that the magnetic field is depleted near the core, hence we call them diamagnetic vortices. In addition, they break $C$ and $P$ symmetries unlike their cousins studied in Ref.~\cite{Anber:2015kxa}. 

The plan of this paper is as follows. In Sec.~\ref{sec:symmetry breaking}, we describe the action of the $\uzua$ Chern Simons gauge theory and sumarize some basic properties of the constant $\ua$ magnetic field solution. The perturbative vacuum structure of the theory is studied in great detail in our accompanying paper \cite{3dsuperconductor}. We study the topological vortex solution by making use of Nielsen-Olesen like Ans\"atze in Sec.~\ref{sec:vortex}. We also obtain the asymptotic behavior of the solution for both the near core and large radius limits and show that the diamagnetic vortices break both $C$ and $P$ symmetries. Next, we numerically integrate the equations of motion to obtain the full solution for various values of the winding number. In Sec.~\ref{sec:properties}, we calculate the flux, charge and energy of the vortices. In Sec.~\ref{sec:bkt}, we sketch the essential elements of the BKT phase transition and derive the critical temperature. We conclude with a discussion of our results in Sec.~\ref{sec:discussion}.     

\section{Spontaneous Symmetry Breaking by $\rm{\bf U(1)}_{\scriptscriptstyle \bf \cal A}$ Magnetic Field}
\label{sec:symmetry breaking}

Before describing the vortex solutions in this theory, we first sumarize our results in Ref.~\cite{3dsuperconductor} for completeness. We consider the $\uzua$ Chern-Simons theory with the action \cite{Anber:2015kxa,3dsuperconductor}
\ba \bald \label{action}
S &= \int d^3x \biggl [-\frac{1}{4} \mathcal{F}_{\mu\nu} \mathcal{F}^{\mu \nu} -\frac{1}{4} \mathcal{Z}_{\mu\nu} \mathcal{Z}^{\mu \nu}+ \mu_1 \epsilon^{\mu \nu \alpha} \mathcal{F}_{\mu\nu} \mathcal{Z}_{\alpha}~~~~~ \\ 
& \hskip 0.5cm + \frac{\mu_2}{2} \epsilon^{\mu \nu \alpha} \mathcal{Z}_{\mu\nu} \mathcal{Z}_{\alpha} + |D_\mu \varphi|^{2} -m^2|\varphi|^{2}  \biggr ] \,,
\label{the main action of the paper}
\eald
\ea
where $\mathcal{F}_{\mu \nu} = \partial_\mu \mathcal{A}_\nu - \partial_\nu \mathcal{A}_\mu$, $\mathcal{Z}_{\mu \nu} = \partial_\mu \mathcal{Z}_\nu - \partial_\nu \mathcal{Z}_\mu$, $D_\mu = \partial_\mu - i e \mathcal{Z}_\mu$ and $m^2>0$. The Chern Simons coupling constants $\mu_1$ and $\mu_2$, and mass parameter $m$ have mass dimension $M$, whereas the gauge coupling constant $e$ has mass dimension $M^{1/2}$. We use natural units $c=1$, $\hbar =1$, set $k_B =1$, $\epsilon^{012} =1$, and use the metric $\eta_{\mu\nu} = {\rm diag}(1, -1,-1)$ in what follows\footnote{In this paper, we drop the $\lambda |\varphi|^4/4$ term in the action as this does not change the physics of the system, but introduces some trivial corrections. For a detailed discussion, see Ref.~\cite{3dsuperconductor}.}.  

The equations of motion read
\ba \bald \label{field equations}
&\partial_\beta \mathcal{F}^{\beta \sigma} + \mu_1 \epsilon^{\beta \alpha \sigma} \mathcal{Z}_{\beta \alpha} = 0 \com \\
&\partial_\beta \mathcal{Z}^{\beta \sigma} + \mu_1 \epsilon^{\beta \alpha \sigma} \mathcal{F}_{\beta \alpha} + \mu_2 \epsilon^{\beta \alpha \sigma} \mathcal{Z}_{\beta \alpha} +j^\sigma=0 \com \\ 
&D_\beta D^{\beta} \varphi +m^2 \varphi=0 \com
\eald 
\ea
where
\be\label{current}
j^{\sigma} = i e \bigl [ \varphi^{*} D^{\sigma} \varphi - (D^{\sigma} \varphi)^{*} \varphi \bigr ] \per
\ee
There exists a solution to the field equations (\ref{field equations}) of the form
\be\label{const soln}
B_{\cal A} = B \com ~~~~~{\cal Z}_b^0(\varphi) = -\frac{\mu_1 B}{e^2 |\varphi|^2} \com
\ee
where $B$ is a constant. An effective potential for the Higgs field in this background can be defined as
\be\label{V_eff B}
V_{\rm eff}(\varphi, B) = \frac{\mu_1^2 B^2}{e^2|\varphi|^2} +m^2 |\varphi|^2 
\per
\ee 
The Higgs potential (\ref{V_eff B}) has  a minimum at 
\be\label{B minima}
\varphi_0 = \pm \sqrt{\frac{\mu_1 |B|}{e m}} \com
\ee
at which the background ${\cal Z}_\mu$ field takes the value
\be \label{Z_0 background}
{\cal Z}_b^0(\varphi_0) = -\frac{m}{e} \frac{|B|}{B} \per
\ee
In what follows, we will take $B>0$ without loss of generality. Thus, even though the Higgs potential does not have a tachyonic mass parameter, the presence of a constant $\ua$ magnetic field  breaks the $\uz$ symmetry by turning on the term $(\mu_1^2 B^2)/(e^2|\varphi|^2)$ in the effective Higgs potential given by \eref{V_eff B}. This suggests that the vacuum manifold of $\varphi$, $\cal{M}$, has non-trivial topology, namely $\pi_1({\cal M}) = {\mathbb Z}$. Therefore, there exists a vortex solution, which we describe in the next section.

\section{Topological Vortex Solution}
\label{sec:vortex}

We are interested in cylindrically symmetric vortex solutions, and thus, we consider a Nielsen-Olesen like Ans\"atze for the Higgs and gauge fields \cite{Anber:2015kxa}
\ba\label{ansatz}
\bald
\varphi &= \varphi_0 f(r) e^{i n \theta}\,,  \quad 
\mathcal{Z}_i = -\epsilon^{i j} x_j \frac{Z(r)}{e r^2} \,, \\
\mathcal{Z}_0 &= e  Z_0 (r) \,,\quad ~~~
\mathcal{A}_i = -\epsilon^{i j} x_j \frac{A(r)}{e r^2}  \,, \\
 \mathcal{A}_0 &= e  A_0 (r) \per 
\eald
\ea
Note that we choose the above Ans\"atze such that all the profile functions $f(r), Z(r),  Z_0(r), A(r)$ and $ A_0(r)$ are dimensionless. Upon plugging \eref{ansatz} into the equations of motion (\ref{field equations}), we obtain the following equations for the profile functions:
\ba
\bald
\label{the equations of profile functs}
&f'' + \frac{f'}{r} - (n-Z)^{2} \frac{f}{r^2} + e^4  Z_0^{2} f - m^2 f = 0 \com~~\\
&Z'' -\frac{Z'}{r} + 2e^2 \varphi_0^2 f^2 (n-Z) - 2e^2 r (\mu_1  A_0' + \mu_2  Z_0') = 0 \com~~~  \\
& Z_0'' +\frac{ Z_0'}{r} - 2e^2 \varphi_0^2 f^2  Z_0 - \frac{2}{e^2 r} (\mu_1 A' +\mu_2 Z') =0 \\
&A'' - \frac{A'}{r} - 2 \mu_1 e^2 r  Z_0' = 0 \com  \\
& A_0'' + \frac{ A_0'}{r} - \frac{2\mu_1}{e^2 r} Z' = 0 \com
\eald
\ea
where $r$ is the dimensionful radius. 
The last two equations can be integrated to yield the first order equations:
\be\label{integrated equations}
A' = 2\mu_1 e^2 r Z_0 + {\cal D}_1 r \com \qquad A_0' = \frac{2\mu_1}{e^2 r}Z + \frac{{\cal D}_2}{r} \per
\ee
We set ${\cal D}_1 = Be + 2 \mu_1 m /e^2$ in order to meet the requirement that the magnetic field very far from the core radius\footnote{For simplicity, we postpone the requirement that the magnetic flux has to be conserved for vortices, and construct the solution using an asymptotic magnetic field $B$ that has to be corrected later using the flux conservation (see Sec.~\ref{subsec:flux conservation}).} is $B$. In addition, the vortex solution must be well-behaved at the core which selects ${\cal D}_2 =0$. 

\subsection{The Near-Core and Asymptotic Behavior}

At small $r$, the profile functions can be expanded in positive powers of $r$ and can be solved order by order to satisfy the equations of motion (\ref{the equations of profile functs}). This way, we can fix all the expansion coefficients except the five parameters $f_1,~ z_2,~ z_{00},~a_2,~ a_{00}$. To the leading order in $r$, we find
\ba
\bald
\label{small r}
f(r) \simeq& f_1 r^{|n|}+{\cal O}(r^3) \com \\
Z(r) \simeq& z_2 r^2 + {\cal O}(r^4) \com \\
Z_0(r) \simeq& z_{00} + \frac{a_2 \mu_1 + z_2 \mu_2}{e^2} r^2+ {\cal O}(r^4) \com \\
A(r) \simeq& a_2 r^2 + {\cal O}(r^4) \com \\
A_0(r) \simeq& a_{00}+\frac{z_2 \mu_1}{e^2} r^2 +{\cal O}(r^4) \per 
\eald
\ea

Similarly, at large $r$, the profile functions can be expanded in inverse powers of $r$ (see Ref.~\cite{Anber:2015kxa}). To the leading order we find:
\ba
\bald
\label{large r}
f^\infty(r) &\simeq f(\infty) - \frac{ \mu_1^2 n^2}{B e (B e + 2m \mu_1)} \frac{1}{r^2} + {\cal O}(1/r^4) \com \\
Z^\infty(r) &\simeq Z(\infty) -\frac{4 m \mu_1 n^2 \left(n \mu_1^2 + m \mu_2 \right)}{(B e+2 m \mu_1)^3} \frac{1}{r^2} + {\cal O}(1/r^4) \com \\
Z_0^\infty(r) &\simeq Z_0(\infty) -\frac{2\mu_1^2 m n^2}{e^2 (B e+2 m \mu_1)^2}\frac{1}{r^2} + {\cal O}(1/r^4) \com ~~~~~\\
A^\infty(r) &\simeq A(\infty) +\frac{2 \mu_1^3 n^2 }{(Be + 2 m \mu_1)^4} \Bigg[ 3 \mu_1^2 m \left(n^2+4\right) \\
&\hskip-1cm + \frac{B^2 e^2}{m}+  6 B e \mu_1 + 4 \mu_2 m^2 n + \frac{8 \mu_1^3 m^2}{B e} 
\Bigg] \frac{1}{r^2} + {\cal O}(1/r^4) \com~~ \\
A_0^\infty(r) &\simeq A_0(\infty)+\frac{4 \mu_1^2 m n^2 \left(n \mu_1^2+ m\mu_2 \right)}{e^2 (B e+2 m \mu_1)^3} \frac{1}{r^2} + {\cal O}(1/r^4)\,, \\
\eald
\ea
and the asymptotic values of the profile functions are 
\ba\label{profile asymptotic}
\bald
&f(\infty) = 1 \com ~~ Z(\infty) = \frac{e^2 \varphi_0^2 n}{e^2 \varphi_0^2 + 2 \mu_1^2} \com ~~ Z_0(\infty) = - \frac{m}{e^2} \com ~~~~~\\
&A(\infty) = {\cal C}_0 + \frac{1}{2} B e r^2 - \frac{4 e \mu_1^4 \varphi_0^2 n^2 }{B (e^2 \varphi_0^2 + 2 \mu_1^2)^2}  \ln \frac{e^2 r}{{\cal C}_1} \com\\
&A_0(\infty) = \frac{2 \mu_1 \varphi_0^2 n}{e^2 \varphi_0^2 + 2 \mu_1^2} \ln \frac{e^2 r}{{\cal C}_2} \com
\eald
\ea
where $\varphi_0$ is given by \eref{B minima}. We note that the asymptotic values of the profile functions $Z(\infty)$ and $A_0(\infty)$ in \eref{profile asymptotic} match those of our previous work \cite{Anber:2015kxa} upon substituting $\varphi_0 = v$, where $v$ is the vacuum expectation value of the $\varphi$ field in the $V(\varphi) = -m^2 \varphi^2+\lambda \varphi^{4}/4$ potential with $m^2>0$.

\subsection{C and P Violation in Diamagnetic Vortices}

We can readily understand most of the key physical properties of these diamagnetic vortices without the full numerical solution that will be discussed in the next section. First of all, we recall that the parity ($P$) and charge conjugation ($C$) operators in $2+1$ D act on the position vector, $x^\mu = \left(x^0, x^1,x^2\right)$,  and fields ${\cal A}^\mu(x^\mu) = \left({\cal A}^0, {\cal A}^1,{\cal A}^2\right) (x^\mu)$ and $\varphi(x^\mu)$ as follows (see, e.g., the appendix of Ref.~\cite{Affleck:1982as}):
\ba\label{P transform}
\bald
P: x^{\mu} &\to \bar x^{\mu}=  \left(x^0, -x^1,x^2\right)\com\\
P: {\cal A}^{\mu} (x^\mu) &\to \bar {\cal A}^{\mu} (\bar x^{\mu}) =  \left({\cal A}^0, -{\cal A}^1,{\cal A}^2\right)(\bar x^{\mu}) \com \\
P: \varphi (x^\mu) &\to \bar \varphi (\bar x^{\mu}) =  \varphi (\bar x^{ \mu}) \com
\eald
\ea
and 
\ba\label{C transform}
\bald
C: x^{\mu} & \to x^\mu \com \\
C: {\cal A}^{\mu} (x^\mu) &\to - {\cal A}^{\mu} (x^\mu)  \com \\
C: \varphi (x^\mu) &\to \varphi^* (x^{ \mu}) \per
\eald
\ea
From the transformation properties given in Eqs.~(\ref{P transform}) and (\ref{C transform}), it is easy to check that magnetic field $B$ is odd under both $C$ and $P$, but even under $CP$. Hence, $B$ is a  pseudoscalar. The electric field ${\bm E} = \left(E^1, E^2 \right)$ transforms under $P$ and $C$ as
\ba
\bald
P: {\bm E} &\to \left(-E^1, E^2 \right) \com\\
C: {\bm E} &\to - {\bm E} \per
\eald
\ea
Under $C$ and $P$, the vortex with winding number $n$ transforms as
\ba
C: \varphi(r, \theta) \to \left(\varphi_0 f_n(r) e^{i n \theta}\right)^* = \varphi_0 f_{n}(r) e^{-i n \theta} \com \\
P: \varphi(r, \theta) \to \varphi_0 f_n(r) e^{i n (-\theta)} = \varphi_0 f_{n}(r) e^{-i n \theta} \com
\ea
where we used \eref{P transform} and the fact that $\theta = \tan^{-1} (x_2/x_1)$. For vorticies that conserve $C$ or $P$, the $C$ or $P$ operations would simply correspond to flipping the sign of the winding number $n$, i.e. sending the vortex to anti-vortex since $f_n=f_{-n}$ in that case. However, from the asymptotic behavior of the diamagnetic vortices given in \eref{large r}, we can see that the profile functions $Z(r)$, $A(r)$, and $A_0(r)$ have both even and odd terms in $n$, e.g., $f_n\neq f_{-n}$ (this is clear for $A^\infty (r)$ at ${\cal O}(1/r^2)$, while for the rest of the profile functions one has to go to higher orders to see it). Therefore, under $C$ and $P$, the vortex does not transform into an antivortex, or vice versa. In other words, the diamagnetic vortex solution breaks the $C$ and $P$ symmetry while preserving $CP$. This is not surprising as this solution only exists in the presence of a constant magnetic field $B$ that is odd under $C$ and $P$. The explicit differences between vortex and antivortex solutions of various winding numbers are shown in Fig.~\ref{Fig:profilen}.

\subsection{Numerical Solution}

\begin{figure}[h]
\includegraphics[width=88mm]{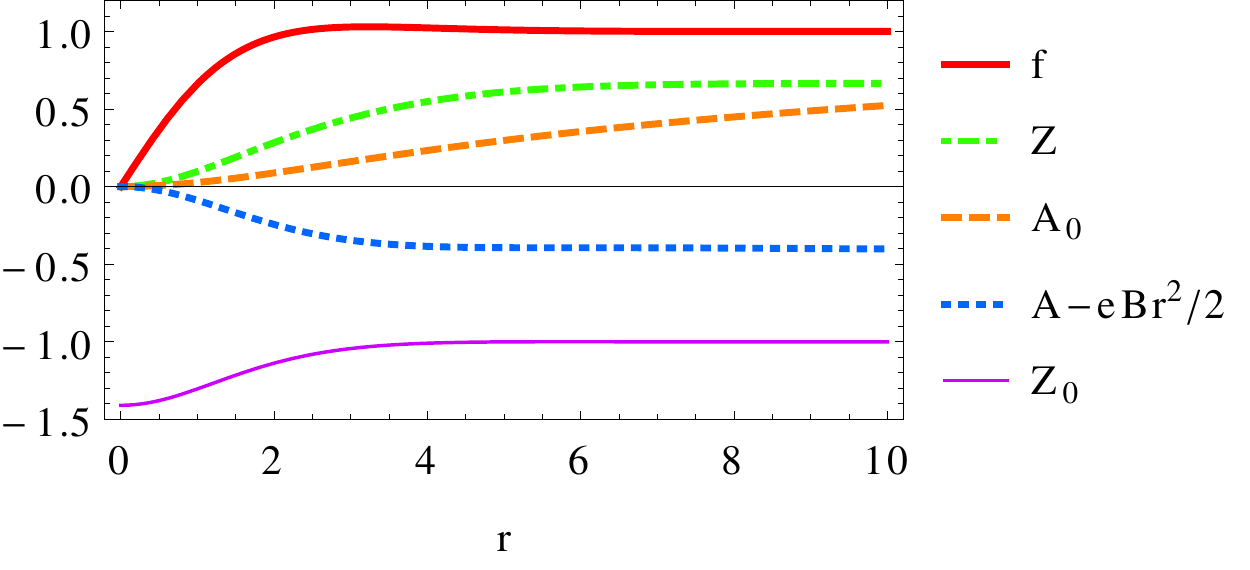}
\caption{Profile functions for $n=1$, $e=1$, $m=1$, $B=1$ and $\mu_1=\mu_2=1/4$ against radial distance $r$ in units of $e^2$.}
\label{Fig:profile}
\end{figure}

\begin{figure}[h]
\begin{center}
\includegraphics[width=88mm]{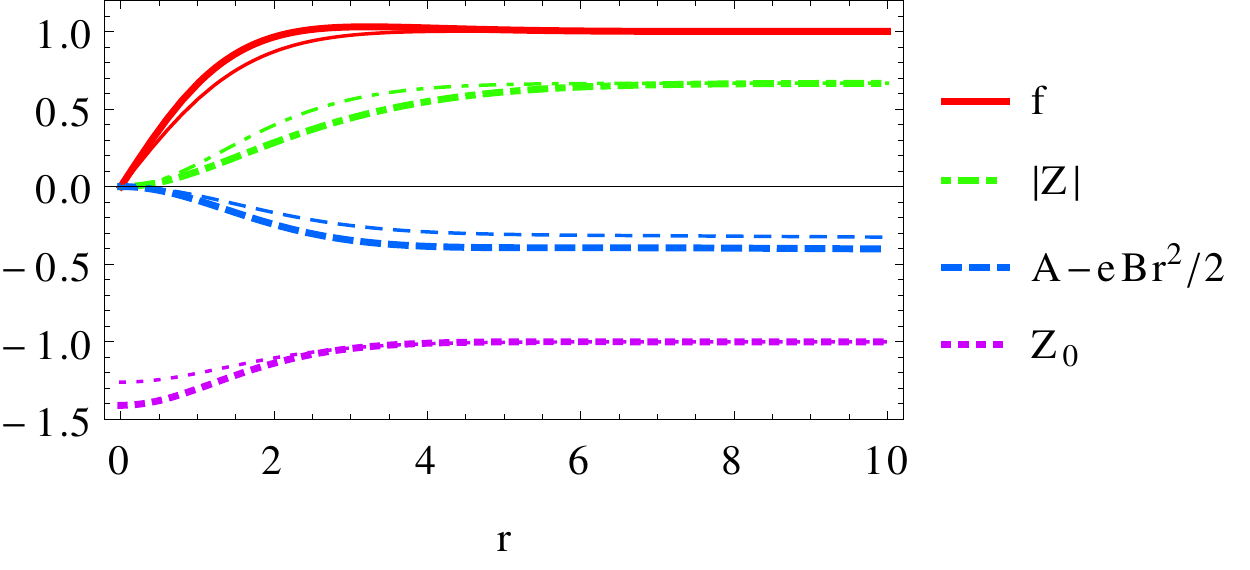}
\includegraphics[width=88mm]{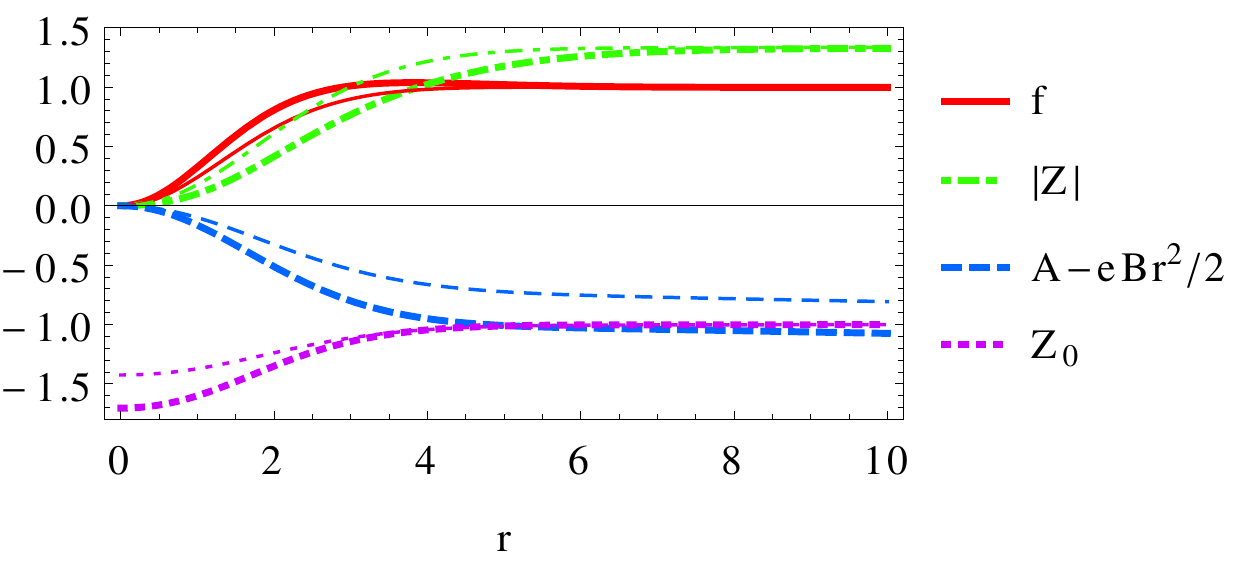}
\includegraphics[width=88mm]{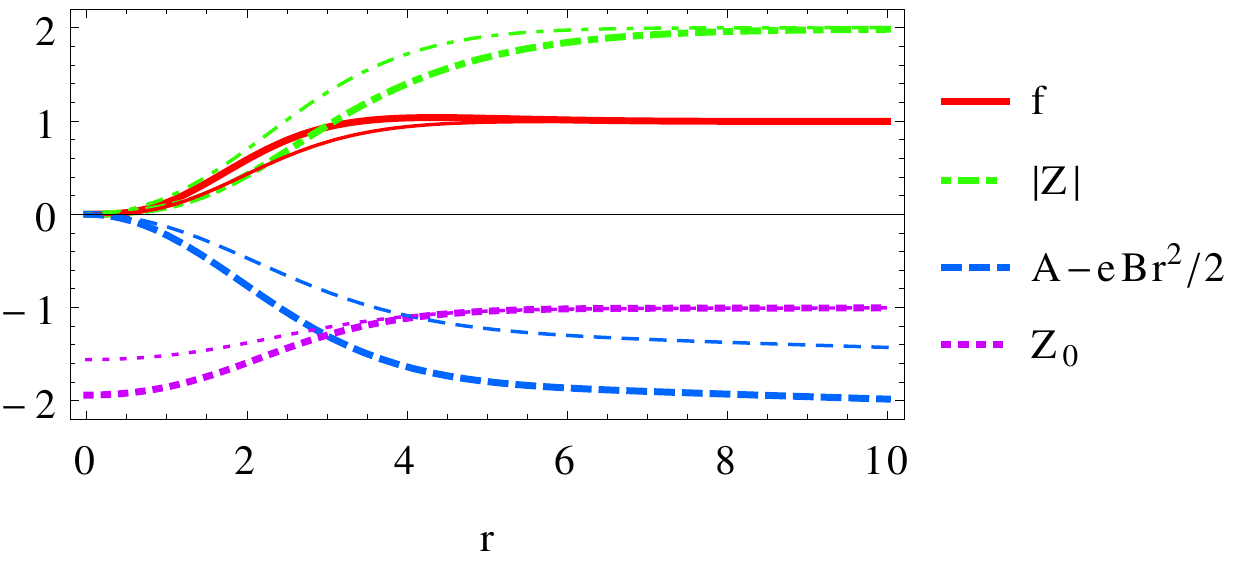}
\caption{Profile functions for $n=\pm1$ (top), $n=\pm 2$ (middle), $n=\pm 3$ (bottom) against radial distance $r$ in units of $e^2$. The profile functions for $n>0$ are shown as thick lines, whereas the profiles for $n<0$ are shown as thin lines. We used the parameters: $e=1$, $m=1$, $B=1$ and $\mu_1=\mu_2=1/4$. Note that $Z$ (and $A_0$ not shown here) are negative for $n<0$. Here we switched the sign of $Z$ for $n<0$ for easy comparison with $n>0$.}
\label{Fig:profilen}
\end{center}
\end{figure}

Numerical results were obtained via a shooting method as implemented in \cite{Burnier:2005he}. For winding number $n=1$ the profile functions are shown in Fig.~\ref{Fig:profile}. 
We see that the profile functions satisfy both the small and large $r$ asymptotic behaviors given in Eqs.~(\ref{small r}) and (\ref{large r}). For instance, at large distance $r$, the Higgs goes to its vacuum expectation value $\varphi_0$, i.e., $f(r)\to 1$, and the ${\cal Z}$ condensate goes to its background value given in \eref{Z_0 background}, i.e., $Z_0(r)\to-m/e$. Because of the magnetic field background (\ref{const soln}), the function $A(r)$ grows like $e Br^2/2$, and thus, we only display $A(r)-e Br^2/2$ that corresponds to the vortex contribution to $A(r)$.
As the solution is neither symmetric or antisymmetric under charge conjugation $\mathcal{C}$ or under parity $P$, the profiles for negative winding numbers are different from the positive winding numbers. We compare diamagnetic vortex solutions with various positive and negative winding numbers $n=\pm1, \pm 2,\pm 3$ in Fig.~\ref{Fig:profilen}.

%

%
%

\begin{figure}[h]
\begin{center}
\includegraphics[width=87mm]{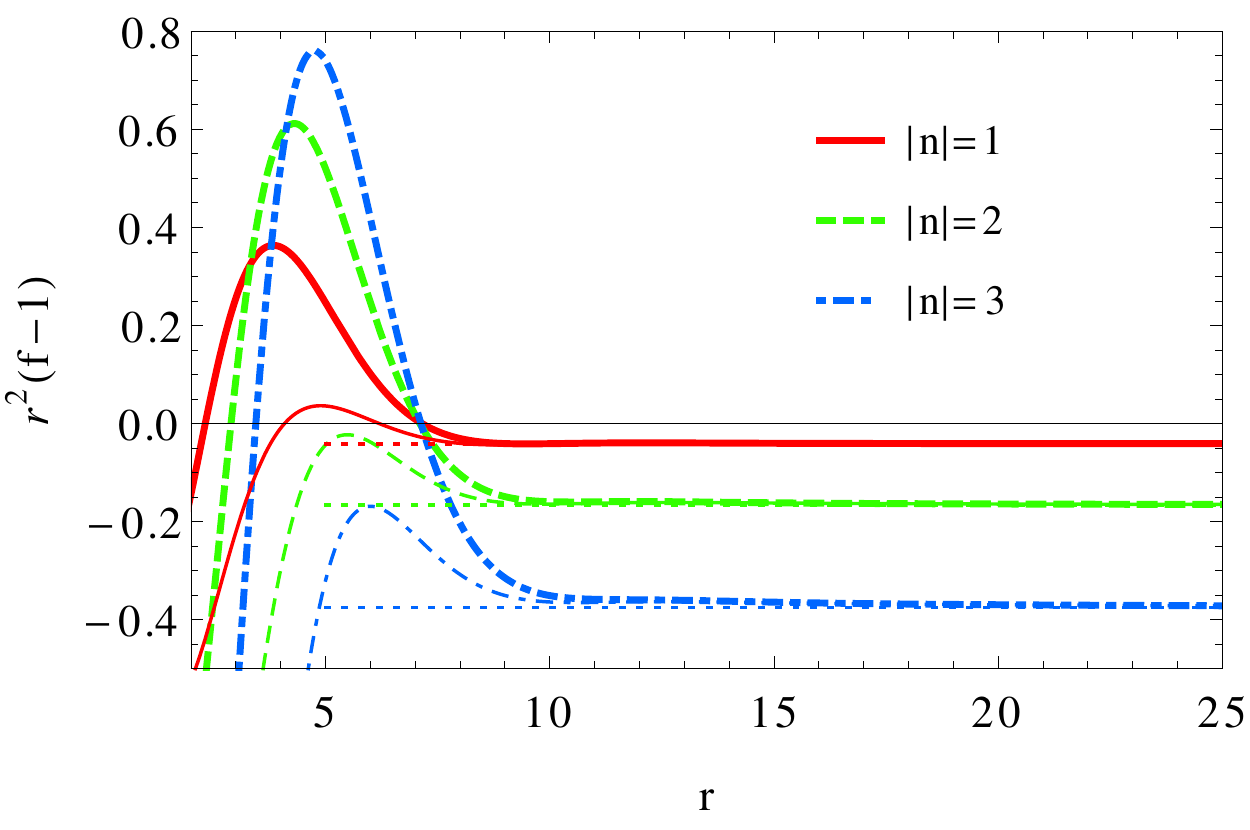}
\includegraphics[width=87mm]{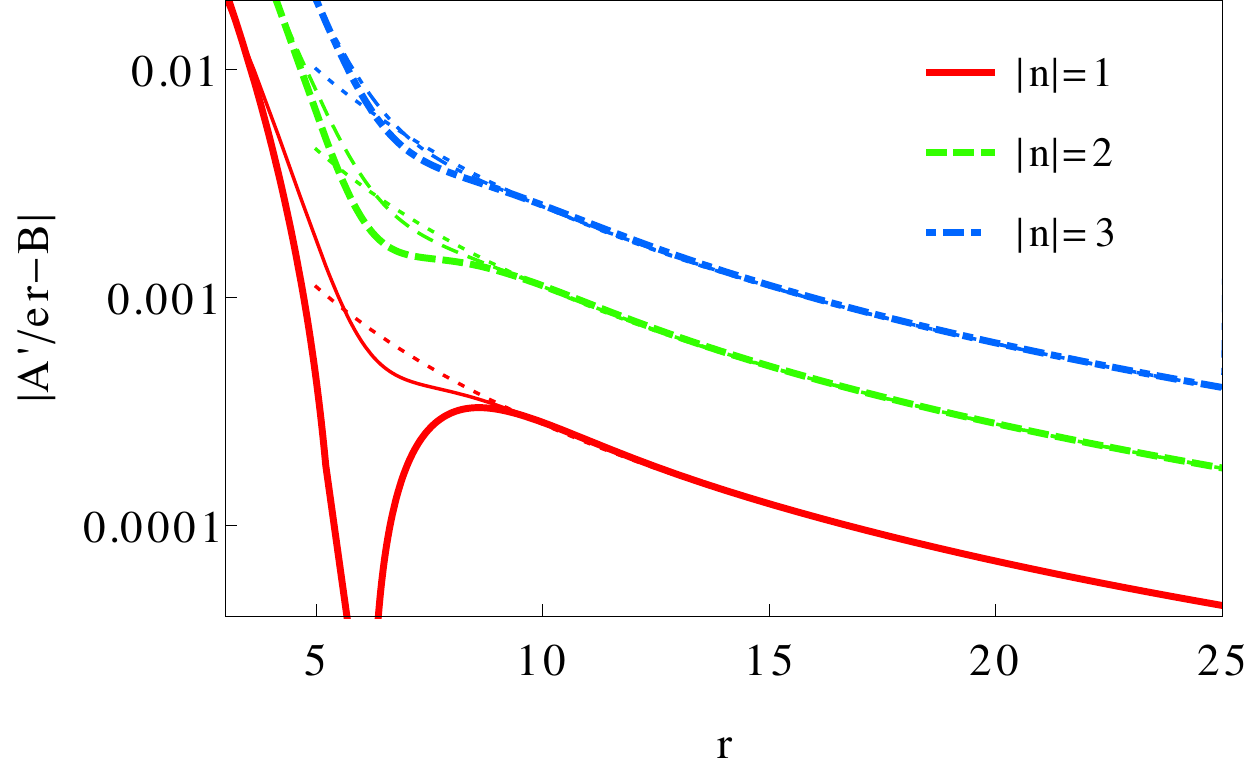}
\caption{Detailed behavior of the scalar profile functions $f$ (top) and $A$ (bottom). Thick (thin) lines correspond to $n>0$ ($n<0$). The thin dotted lines underline the asymptotic behavior derived in \eref{large r}. At large $r$, $1-f(r)$  goes to zero as $1/r^2$, i.e., $r^2[1-f(r)]$ goes to a constant (top). The magnetic field also converges towards its asymptotic value as $1/r^2$ (bottom).}
\label{Fig:large_r2}
\end{center}
\end{figure}

We also provide more precise checks for the asymptotic behavior of the profile functions $f(r)$ and $A(r)$ in Fig.~\ref{Fig:large_r2}. Note that unlike the other known vortex solutions in similar models where the profile function $f$ is monotonous, $f$ first overshoots its asymptotic value $1$, and then turns around again and finally reaches $1$ from below.
%

\section{Physical Properties of Diamagnetic Vortices}
\label{sec:properties}

\begin{figure}[h]
\begin{center}
\includegraphics[width=76mm]{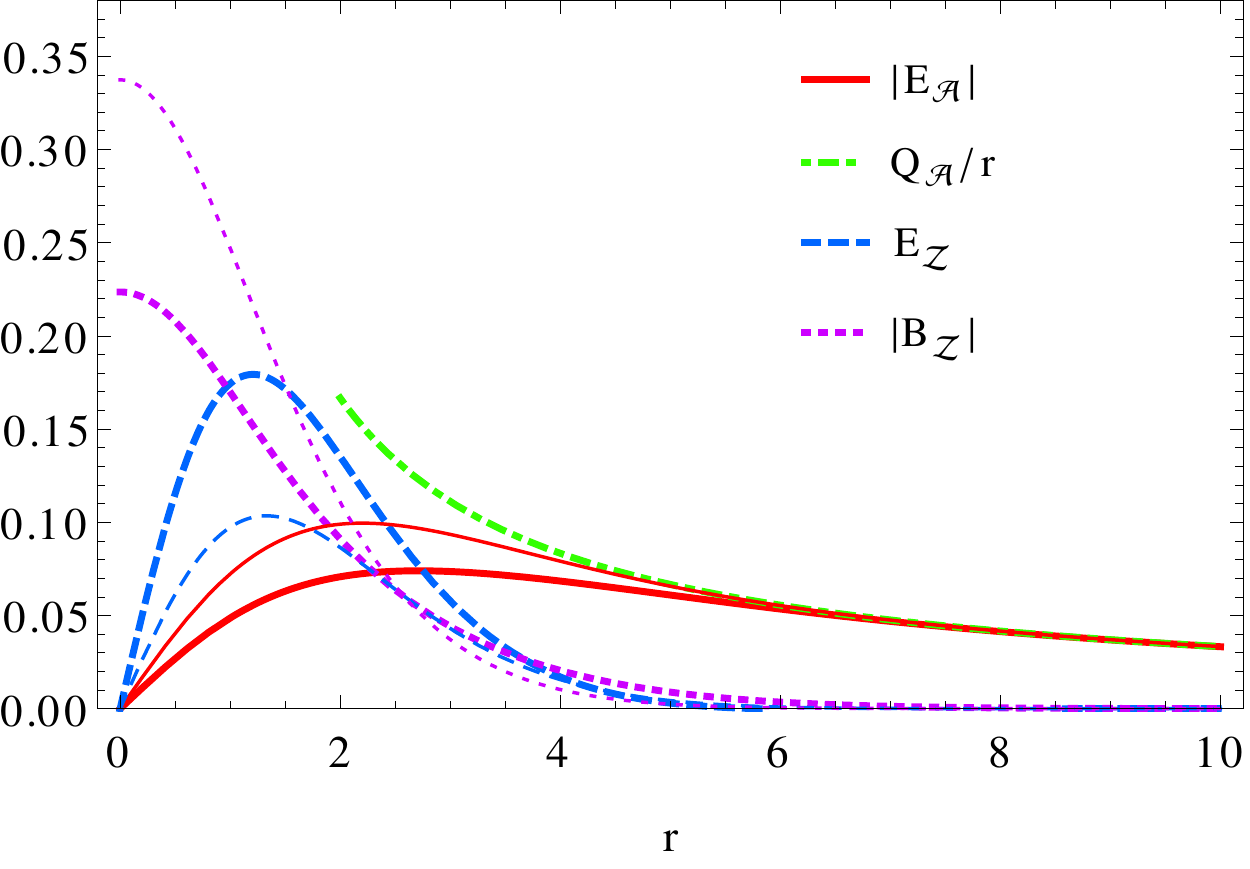}
\includegraphics[width=76mm]{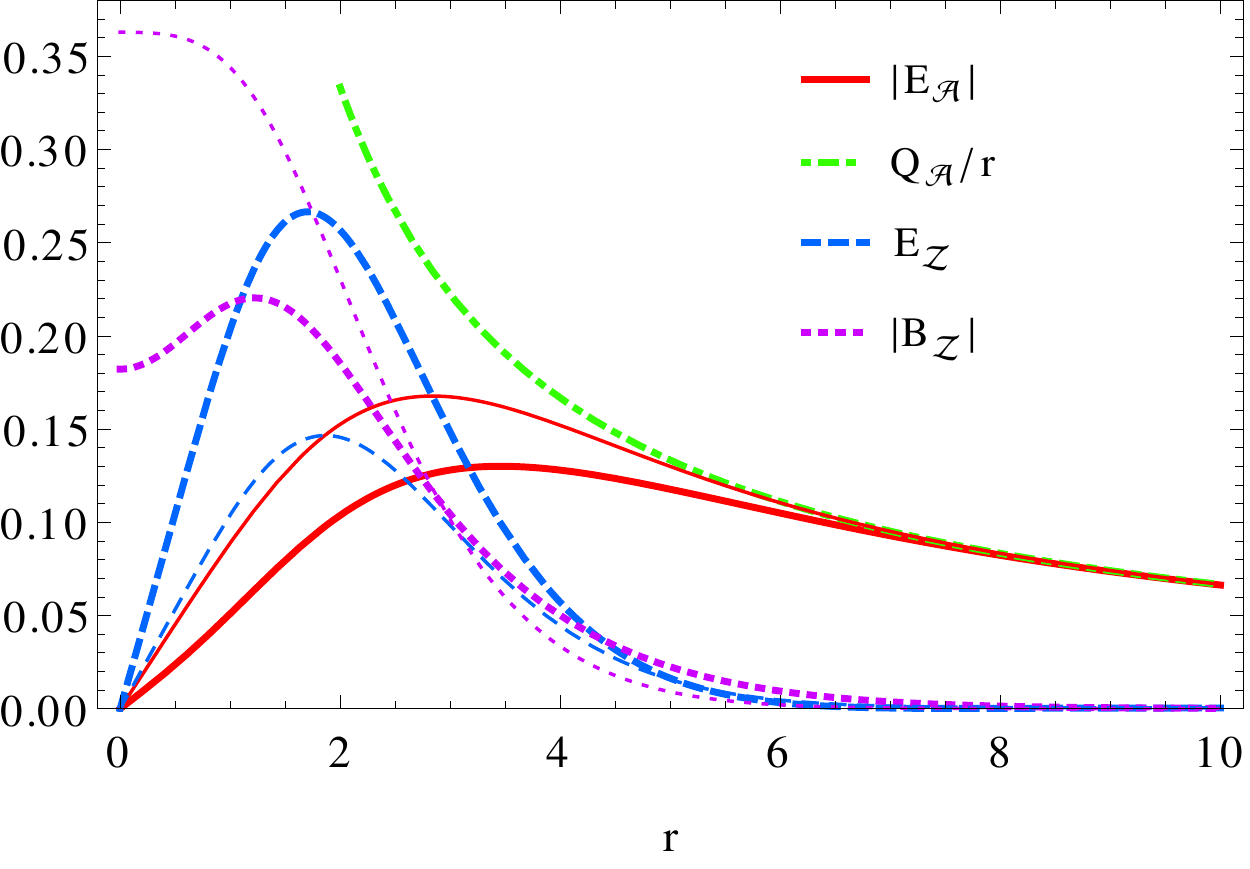}
\includegraphics[width=76mm]{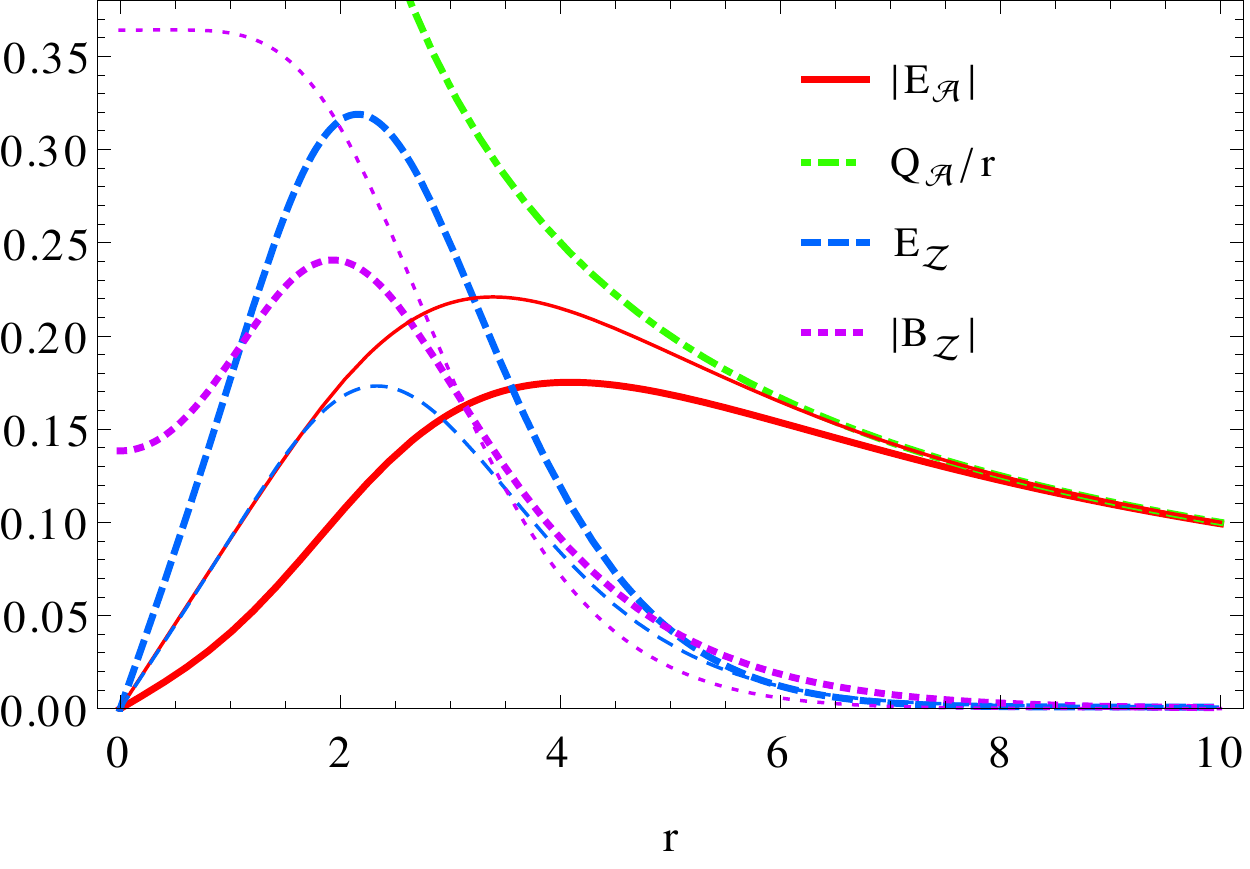}
\caption{Electric and magnetic fields ($B_\mathcal{A}$ is shown in Fig.~\ref{Fig:fields2}). The plots in the top, middle and bottom panels correspond to $n=\pm1$, $n=\pm2$ and $n=\pm 3$, respectively. Positive (negative) winding numbers are shown as thick (thin) lines. The total $\ua$ charge of the vortex is $Q_\mathcal{A}=2 B \mu_1 n/(B e^2 + 2 m e \mu_1)$ from \eref{A charge}, and we can see that $E_\mathcal{A}$ electric field asymptotically reaches $Q_\mathcal{A}/r$ at large $r$. Note also that for $n<0$, $E_\mathcal{A}$ and $B_\mathcal{Z}$ are negative. We show the absolute value of these fields for visually clear comparison. }
\label{Fig:fields}
\end{center}
\end{figure}

\begin{figure}[]
\begin{center}
\includegraphics[width=82mm]{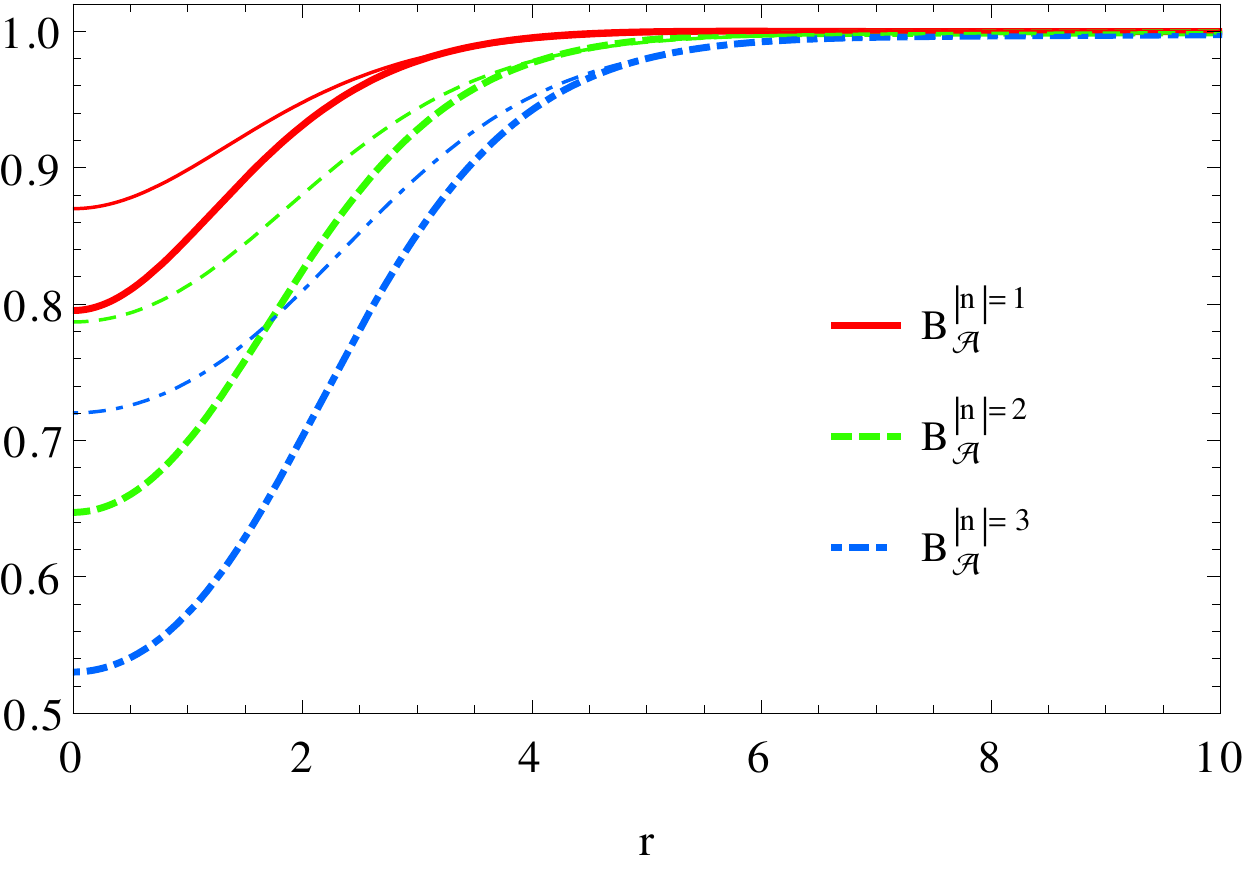}
\caption{Magnetic field $B_\mathcal{A}$ for different winding numbers $n$. The radial dependence of $B_{\cal A}$ is shown as thick (thin) lines for positive (negative) $n$. At infinity $B_\mathcal{A}$ reaches $1$ monotonously, satisfying our assumptions of a constant magnetic field background. In the core of the vortex the magnetic field is depleted for all values of $n$.}
\label{Fig:fields2}
\end{center}
\end{figure}

Using the profile functions given by \eref{ansatz}, we can calculate the electric and magnetic fields of the diamagnetic vortices as:
\ba \label{elmag}
\bald
E_\mathcal{Z} &= e  Z_0' \com \qquad B_{\mathcal{Z}} = \frac{1}{2} \epsilon^{0ij}Z_{ij} = \frac{Z'}{e r} \com \\
E_\mathcal{A} &= e  A_0' \com \qquad B_{\mathcal{A}} = \frac{1}{2} \epsilon^{0ij}F_{ij} = \frac{A'}{e r} \com
\eald
\ea 
$\mathcal{Z}_{0} = e Z_0$ and the kinetic term for the Higgs field can be similarly found as
\be\label{D phi}
|{\bf D} \varphi|^2 = \varphi_0^2 f'^2 + \varphi_0^2 (n-Z)^2 \frac{f^2}{r^2} \per
\ee
The electric and magnetic fields are shown in Figs.~\ref{Fig:fields} and \ref{Fig:fields2}.
%

\subsection{Flux, Charge and Energy}

We calculated all the physical properties of similar vortices with the potential $V(\varphi) = -m^2 \varphi^2+\lambda \varphi^{4}/4$ in Ref.~\cite{Anber:2015kxa}. It is a straightforward exercise to obtain the flux, charge and energy of the diamagnetic vortices.
The ${\cal Z}_\mu$ magnetic flux of a diamagnetic vortex is
\ba \label{phiBz}
\Phi_{B_\mathcal{Z}}^{\rm v} = \frac{2\pi}{e} Z(\infty) 
= \frac{2 \pi n}{e} \frac{ e^2 \varphi_0^2 }{ e^2 \varphi_0^2+2\mu_1^2} \,.
\ea
Note that for the creation of a pair of a vortex and antivortex, the ${\cal Z}_\mu$ flux is automatically conserved as it is proportional to the winding number $n$.
 
\begin{figure}[]
\begin{center}
\includegraphics[width=87mm]{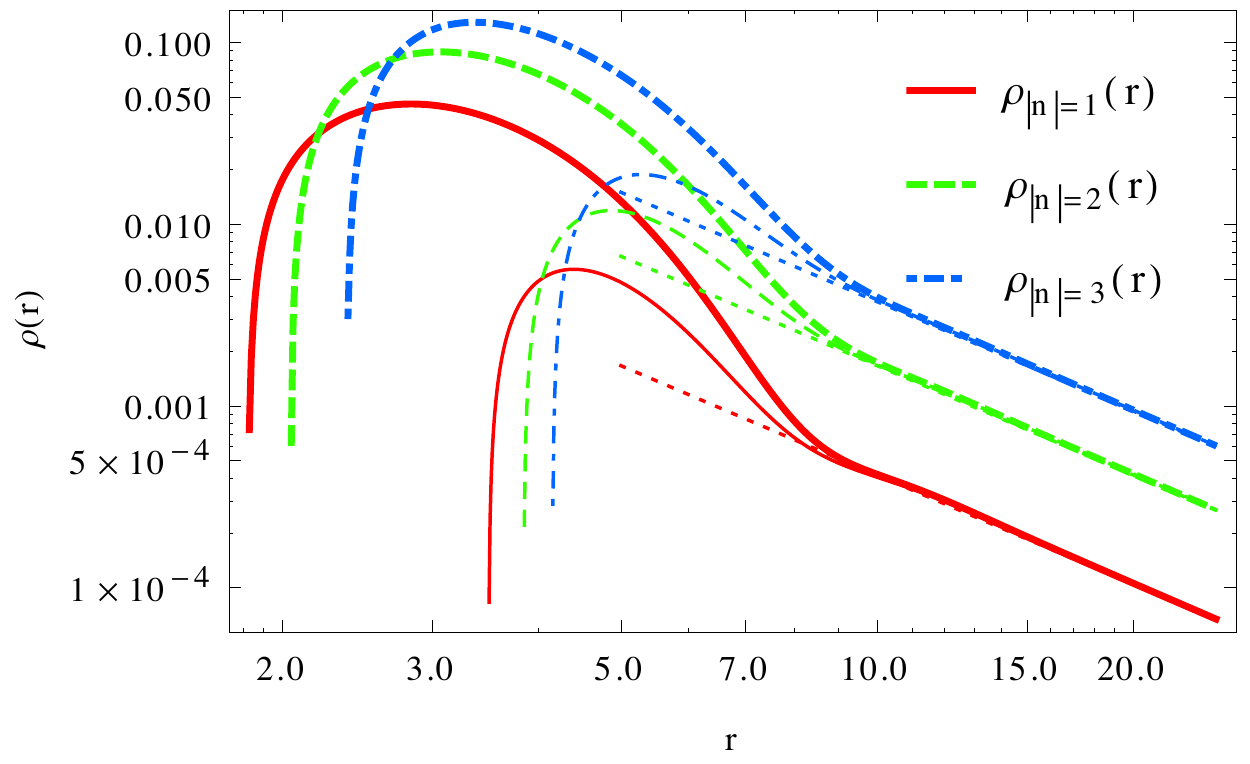}
\caption{The energy density of vortices (thick lines) and anti-vortices (thin lines) for $n=\pm 1\com \pm 2\com \pm 3$. It is clear that vortices and anti-vortices have different core energies, which is expected since the vortex solution breaks the $C$ and $P$ symmetries.}
\label{core energy}
\end{center}
\end{figure}

The long range $\ua$ charge can be found from Gauss's law
\ba \label{A charge}
Q_\mathcal{A} &=& \oint_{S^{1}_{\infty}} d{\bm \ell} \cdot {\bm E}_{\cal A} = 2\pi r e A_0' \nn \\
&=& \frac{4 \pi n}{e}\frac{e^2 \varphi_0^2 \mu_1}{ e^2 \varphi_0^2+2\mu_1^2} \per
\ea
The Hamiltonian density  is given by
\ba \bald \label{H}
\mathcal{H} &= \frac{1}{2} \left( E_{\mathcal{Z}}^{2} + B_{\mathcal{Z}}^{2} + E_{\mathcal{A}}^{2} + B_{\mathcal{A}}^{2} \right) + e^4  Z_0^2 |\varphi|^2 \\
&\hskip 0.5cm + |{\bf D} \varphi|^2 + m^2 |\varphi|^2 \com
\eald
\ea
which can be expressed in terms of the profile functions using Eqs.~(\ref{elmag}) and (\ref{D phi}) as
\ba\label{hamiltonian}
\bald
\mathcal{H} =&\frac{1}{2}\biggl [ e^2  Z_0'^2 + \frac{Z'^2}{e^2 r^2} + e^2  A_0'^2 +\frac{A'^2}{e^2 r^2} + 2 e^4 \varphi_0^2  Z_0^2 f^2~~~~~~~\\
& +2 v^2 (n-Z)^2 \frac{f^2}{r^2} + 2 v^2 f'^2 +2 m^2 \varphi_0^2 f^2  \biggr] \per
\eald
\ea
Integrating each term over $\mathbb{R}^2$, the total energy of a vortex reads
\ba \label{Ev}
\mathcal{E} = \mathcal{E}_{\rm c}+\Eir  \,,
\ea
where $\mathcal{E}_{\rm c}$ is the vortex core energy that we compute numerically, as shown in Fig.~\ref{core energy}, and $\Eir$ is the infrared part of the energy. 
Imposing an IR cutoff at radial distance $r = L$ we find
\ba \label{vortex energy}
\Eir&=& 2\pi \left(\frac{1}{4} B^2 + \frac{B \mu_1 m}{e}\right) L^2\notag\\&&+\frac{4 \pi \mu_1^2 n^2}{e^2} \frac{Be - 2 m \mu_1}{Be + 2 m \mu_1} \ln \frac{L}{r_{\rm c}}  \com
\ea
where the first term is the IR contribution from the background magnetic field and ${\cal Z}_0$ condensate. Note also that the last term suggests that energy of a vortex has a negative contribution. This might seem puzzling at first as negative energy would be naively associated with some sort of instability in the ground state of the system. However, we emphasize at this point that we have not made use of the magnetic flux conservation yet, and thus, the magnetic field $B$ is not the {\it true} background field. Next, we calculate the value of the asymptotic magnetic field value, which upon substituting in \eref{vortex energy} yields a finite and positive vortex energy.

%

\subsection{Conservation of Magnetic Flux}
\label{subsec:flux conservation}

For a self consistent vortex solution, we also need to take into account the fact that the magnetic flux is a conserved quantity. Using the Bianchi identity
\ba
\bald
\epsilon^{\alpha \mu\nu} \partial_\alpha {\cal F}_{\mu\nu} &=0 \com \\
\eald
\ea
integrating over a surface ${\mathbb R}^2$, and making use of Stokes' theorem, we obtain
\ba
\bald
\frac{\partial \Phi_{B_{\cal A}}}{\partial t} &= - \oint d {\bm \ell} \cdot {\bm E}_{\cal A} \com \\
\eald
\ea
where
\ba
\Phi_{B_{\cal A}}  = \int_{{\mathbb R}^2} dS~B_{\cal A} \com
\ea 
is the magnetic flux, and ${\bm E_{\cal A}}$ is the electric field. Thus, assuming a manifold with electric fields perpendicular to its boundary, i.e., $\oint d {\bm \ell} \cdot {\bm E}_{\cal A}=0$, the magnetic flux must be conserved. The flux conservation also applies to all compact $2$ dimensional manifolds without boundaries.

Suppose that we start with a constant background magnetic field $B_{\cal A} = \bar B$ on ${\mathbb R}^2$ whose flux is equal to
\ba \label{flux bar B}
\Phi_{B_{\cal A}} = \int_0^{2\pi} d\phi \int_{0}^{L} dr~r~ \bar B = \pi L^2 \bar B \com
\ea
where $L$ is an IR cutoff characterizing the system size and $r_c$ is the core radius\footnote{Notice that there is also core magnetic flux which should be determined numerically but it can be absorbed in the redefinition of $r_c$.}.  As we can see from Fig.~\ref{Fig:fields2}, the magnetic field strength decreases near the vortices, hence the name diamagnetic. The magnetic flux of a vortex can be obtained analytically from the asymptotic value of the ${\cal A}_\mu$ field in \eref{profile asymptotic}:
\ba \label{flux B}
\Phi_{B_{\cal A}}  &=& \int_0^{2\pi} d\theta \int_{0}^{L} dr~r~ \frac{A'}{er} \nn \\
&=& \pi L^2 B - \frac{8 \pi \mu_1^3 m n^2 }{e (Be + 2 \mu_1 m)^2}  \ln \frac{L}{r_c} + {\cal O} (1/L) \per~~~
\ea
Requiring that the total magnetic flux in the system upon creating a vortex or antivortex is conserved, and noting that the logarithmically divergent piece is negative for both the vortex and antivortex, we have a relation between the asymptotic value of the vortex magnetic field $B$ and the {\it initial} background magnetic field $\bar B$:
\ba
\bar B = B - \frac{8 \mu_1^3 m n^2 }{e (Be + 2 \mu_1 m)^2} \frac{\ln L/r_c}{L^2} \per
\ea
Solving $B$ in terms of $\bar B$ to leading order in inverse powers of $L$, we obtain
\ba \label{bar B}
B \simeq \bar B + \frac{8 \mu_1^3 m n^2 }{e (\bar Be + 2 \mu_1 m)^2} \frac{\ln L/r_c}{L^2} \per
\ea
Since our diamagnetic vortices lower the magnetic field $B_{\cal A}$ in their core [see \eref{profile asymptotic} and Fig.~\ref{Fig:fields2}], the background magnetic field has to increase from $\bar B$ to $ B$ given by \eref{bar B} as required by magnetic flux conservation. This correction will be crucial for the pair of a vortex and an antivortex to have a finite positive energy as we show in the next section (however, notice also that $\bar B\rightarrow B$ as $L \rightarrow \infty$.).

Next, we impose the condition of flux conservation by using \eref{bar B} to find the IR energy
\ba \nonumber
\Eir 
&\approx& 2\pi \left(\frac{1}{4} \bar B^2 + \frac{\bar B \mu_1 m}{e}\right) L^2 \notag\\&&+\frac{4 \pi \mu_1^2 n^2}{e^2} \frac{\bar B e}{\bar B e + 2 m \mu_1} \ln \frac{L}{r_{\rm c}}\,.  \label{energy ir}
\ea
Thus, by subtracting the background energy we obtain the IR energy of a single vortex:
\ba 
\Eir^{\rm v} \approx\frac{4 \pi \mu_1^2 n^2}{e^2} \frac{\bar B e}{\bar B e + 2 m \mu_1} \ln \frac{L}{r_{\rm c}}\, ,
\ea
which is now positive.

\subsection{Interaction between Two Diamagnetic Vortices}

The dynamics of a vortex-antivortex system can be studied by calculating the interaction energy, similar to what we have done in Ref.~\cite{Anber:2015kxa}. To this end, we use the following approximate field configurations for the Higgs and gauge fields assuming that the two vortices are separated by a distance $R$ larger than the core radii of the pair, i.e., $R \gg r_c$:
\ba \bald \label{2vortex}
&\varphi \cong  \varphi_0  e^{i n \theta_1({\bf x}- {\bf x_1}) - i n \theta_2({\bf x}-{\bf x_2})} \\ &\quad\quad\quad\times \left[1+ \frac{f_2}{|{\bf x} - {\bf x_1}|^2} + \frac{f_2}{|{\bf x} - {\bf x_1}|^2} \right]\com  \\
&\mathcal{Z}_0 \cong  -\frac{m}{e} + \frac{\bar z_2}{|{\bf x} - {\bf x_1}|^2} + \frac{\bar z_2}{|{\bf x} - {\bf x_1}|^2} \com \\
&\mathcal{Z}_i \cong -\frac{Q_\mathcal{A}}{4 \pi \mu_1} \epsilon_{i j} \left[ \frac{({\bf x} - {\bf x_1})_j}{|{\bf x} - {\bf x_1}|^2} - \frac{({\bf x} - {\bf x_2})_j}{|{\bf x} - {\bf x_2}|^2}  \right] \com \\
&\mathcal{A}_0 \cong \frac{Q_\mathcal{A}}{2\pi} {\rm ln} \frac{|{\bf x} - {\bf x_1}|}{|{\bf x} - {\bf x_2}|} \com  \\
&\mathcal{A}_i \cong - \epsilon_{i j} \left[ \frac{n \mathcal{C}_2}{e} + a_L \ln \frac{|{\bf x} - {\bf x_1}|}{r_c}  \right] \frac{({\bf x} - {\bf x_1})_j}{|{\bf x} - {\bf x_1}|^2} \\
&- \epsilon_{i j} \left[ -\frac{n \mathcal{C}_2}{e} + a_L \ln \frac{|{\bf x} - {\bf x_2}|}{r_c}  \right] \frac{({\bf x} - {\bf x_2})_j}{|{\bf x} - {\bf x_2}|^2} - \epsilon_{i j} {\bf x}_j \frac{1}{2} B\,,
\eald
\ea
where 
\ba
\bald
f_2 &= - \frac{ \mu_1^2 n^2}{B e (B e + 2m \mu_1)}\,,\\
\bar z_2 &= -\frac{2\mu_1^2 m n^2}{e (B e+2 m \mu_1)^2}\,,\\
a_L &=  - \frac{4 \mu_1^4 \varphi_0^2 n^2 }{B (e^2 \varphi_0^2 + 2 \mu_1^2)^2} \per 
\eald
\ea
Substituting the above expressions in Eq. (\ref{hamiltonian}), imposing the condition of magnetic flux conservation and using $B \rightarrow \bar B$ from \eref{bar B} for a pair of a vortex and an antivortex [note that this doubles the logarithmic term in \eref{bar B}], we obtain the total energy of a pair of interacting vortices:
\begin{eqnarray}
\Eir^{\rm pair} &\approx& 
\frac{8 \pi \mu_1^2 n^2}{e^2} \frac{\bar B e}{\bar B e + 2 m \mu_1} \ln \frac{R}{r_{\rm c}}\,.
\end{eqnarray}
As was explained in Ref.~\cite{Anber:2015kxa}, the energy receives contributions from two parts: the electrostatic energy of the long range $U(1)_{\cal A}$ field and Goldstone background.

\section{Berezinsky-Kosterlitz-Thouless Transition}
\label{sec:bkt}

In the previous sections, we studied the diamagnetic vortices appearing in $\uzua$ Chern-Simons theory in the background of a constant $\ua$ magnetic field  and ${\cal Z}_0$ condensate. Since these vortices are heavy (they cost core energy) they can not alter the low-energy description of our system which is simply a massless $\ua$ theory. 
At any finite temperature,  these vortices are suppressed by a Boltzmann factor  $e^{- \mathcal{E}_{\rm c}/T}$, where $\mathcal{E}_{\rm c}$ is the core energy. Hence, it is  expensive to produce them at zero or low temperature. However, at some critical temperature, $T_c$, it will be entropically favored to produce them, and hence they will proliferate. Since these vortices are charged under $\ua$, their proliferation will break the later group. This is the celebrated Berezinsky-Kosterlitz-Thouless (BKT) transition \cite{Berezinsky:1970fr,Kosterlitz:1973xp}, which can be understood by studying the partition function of a Coulomb gas of vortices. 

To this end, let us consider a gas of positively and negatively charged objects interacting via a Coulomb-like potential. The grand partition function of the gas is given by
\begin{eqnarray}\nonumber
Z&=&\sum_{N_+,N_-=0}^\infty\frac{\left(\frac{\xi_+}{\varepsilon^2}\right)^{ N_+}\left(\frac{\xi_-}{\varepsilon^2}\right)^{ N_-}}{N_+!N_-!}\int \prod_{i=0}^{N_+} d^2 r^a_i\int \prod_{j=0}^{N_-} d^2 r_j^b\\
\nonumber
&&\times \exp\left[\kappa^2\sum_{i\geq j} \log\frac{| \pmb r^a_i- \pmb r^a_j|}{\varepsilon}+\kappa^2\sum_{i\geq j} \log\frac{| \pmb r^b_i- \pmb r^b_j|}{\varepsilon}\right.\\
&&\left.-\kappa^2\sum_{i\geq j} \log\frac{| \pmb r^a_i- \pmb r^b_j|}{\varepsilon} \right]\,,
\end{eqnarray}
where
\begin{eqnarray}
\kappa^2=\frac{8\pi \mu_1^2}{e^2 T}\frac{\bar Be}{\bar Be+2m\mu_1}\,,
\end{eqnarray}
$\varepsilon$ is the UV cutoff length, and $\xi_+$ and $\xi_-$ are the fugacities (recall that the fugacity is $e^{-\mathcal{E}_{\rm c}/T}$) of the positive and negative charges, respectively. The coordinates $\pmb r^a$ are used to label the positively charged objects, while $\pmb r^b$ are used for the negative ones. Next, we impose the charge neutrality on the two dimensional system by demanding that $N_+=N_-=N$. Hence, the factor  $\xi_+^{ N_+}\xi_-^{ N_-}$ becomes $\left(\xi_+\xi_-\right)^N\equiv \xi^{2N}$, where we defined $\xi^2\equiv \xi_+\xi_-$. Thus, the partition function reduces to
\begin{eqnarray}\nonumber
Z&=&\sum_{q=\pm1}\sum_{N=0}^\infty\frac{\left(\frac{\xi}{\varepsilon^2}\right)^{2N}} 
{(N!)^2}\int \prod_{i=0}^{2N} d^2 r_i\\
\label{resulting Z}
&&\times \exp\left[\kappa^2\sum_{i\geq j} q_iq_j \log\frac{| \pmb r_i- \pmb r_j|}{\varepsilon}\right]\,,
\end{eqnarray}
such that only neutral configurations of charges are used in computing $Z$. 
Then, one uses the partition function (\ref{resulting Z}) to derive renormalization group equations (RGE)s for $\xi$ and $\kappa$. The RGE for $\xi$ can be obtained by considering  a single neutral pair of charges \cite{KardarBook}:
\begin{eqnarray}\nonumber
Z^{(1)}&=&1+\frac{\xi^2}{\varepsilon^4}\int d^2r_1 d^2r_2 e^{-\kappa^2\log\frac{|\pmb r_1-\pmb r_2|}{\varepsilon}}\\
\label{single pair in Z}
&=&1+\xi^2(\varepsilon)\left(\frac{L}{\varepsilon}\right)^{4-\kappa^2}\,,
\end{eqnarray}
where $L$ is the system size, and $\xi^2(\varepsilon)$ indicates that the fugacity is being calculated given the UV cutoff $\varepsilon$. The RG invariance of the contribution (\ref{single pair in Z}) to the partition function demands that
\begin{eqnarray}\label{rescaling}
\xi^2(\varepsilon)\left(\frac{L}{\varepsilon}\right)^{4-\kappa^2}=\xi^2\left(e^b\varepsilon\right)\left(\frac{L}{e^b\varepsilon}\right)^{4-\kappa^2}
\end{eqnarray}
under the rescaling $\varepsilon\rightarrow e^b\varepsilon$. Differentiating (\ref{rescaling}) with respect to $b$ at $b=0$, we obtain the RG equation for $\xi$:
\begin{eqnarray}
\frac{d\xi}{db}=\left(2-\frac{\kappa^2}{2}\right)\xi\,.
\end{eqnarray}
This equation tells us that the fugacity remains irrelevant at large $\kappa$ or at low temperature. However, at small values of $\kappa$ or at high temperature the fugacity becomes relevant: the vortices proliferate indicating a phase transition. This is the BKT phase transition which takes place at $\kappa=2$ or
\begin{eqnarray}
T_c=\frac{2\pi \mu_1^2 \bar B}{\bar B e^2+2m e\mu_1}\,.
\end{eqnarray}
%

\section{Discussion}
\label{sec:discussion}

In this work, we studied the non-perturbative sector of the $\uzua$ Chern Simons gauge theory in the background of $\ua$ magnetic field, whose detailed structure is described in our accompanying paper \cite{3dsuperconductor}. This theory admits topological vortex solutions with novel properties that we investigated by analytic and numerical techniques. 

First of all, these vortices exhibit long range interactions since they are charged under the unbroken $\ua$. Besides,  they deplete the $\ua$ magnetic field near their cores. This can be understood by noting that the $\ua$ gauge field ${\cal A}_\mu$ becomes topologically massive in the near core region. In this regard, they behave as a diamagnetic material. As was discussed in Sec.~\ref{sec:properties}, the vortex solution breaks and $C$ and $P$ symmetries, while preserving $CP$. This is also not surprising as the background magnetic field $B$ is odd under $C$ and $P$ and even under $CP$.

As is shown in our work \cite{3dsuperconductor}, the $\uzua$ Chern Simons gauge theory exhibits superconductivity. The role of the condensate is played by the complex scalar field $\varphi$ which develops a vacuum expectation value because of the external $\ua$ magnetic field, and thus breaks the $\uz$ symmetry spontaneously. Naively, one would expect that the superconductivity may be ruined at finite temperature because of symmetry restoration of the $\uz$ perturbatively. However, we showed in Ref.~\cite{3dsuperconductor} that this is not the case. Furthermore, the superconducting vacuum might be in danger because of the non-perturbative sector of the theory, namely the proliferation of the diamagnetic vortices. In Sec.~\ref{sec:bkt}, we studied the BKT phase transition due to these vortices, and we found that the critical temperature is $T_c= 2\pi \mu_1^2 \bar B/(\bar B e^2+2m e\mu_1)$. For temperatures $T > T_c$, diamagnetic vortices proliferate resulting in a more complicated ground state. However, the BKT transition can be postponed by taking $\mu_1 \gg \bar B e/2m$. In this limit $T_c \simeq \pi \mu_1 \bar B/ (2m e)$. Hence by increasing $\bar B$ and keeping the hierarchy $\mu_1 \gg \bar B e/2m$, we can postpone the BKT phase transition to arbitrarily high temperatures. In this regard, we can achieve superconductivity at all temperatures \cite{3dsuperconductor}. 

\acknowledgements

This work was supported by the Swiss National Science Foundation. Y.B. is supported by the grant PZ00P2-142524.

\bibliographystyle{apsrev4-1}

\end{document}